# AN RFQ INJECTION SYSTEM FOR THE HRIBF

Y. Zhang and P. E. Mueller

Oak Ridge National Laboratory, P.O. Box 2008, Oak Ridge, TN 37831-6368, USA

## ABSTRACT

At the Holifield Radioactive Ion Beam Facility (HRIBF) at Oak Ridge National Laboratory (ORNL), molecular ions extracted from a positive ion source and subsequently broken up in a charge-exchange cell produce Radioactive Ion Beams (RIBs) with several hundred eV energy spread, preventing effective magnetic isobar separation. In order to perform magnetic isobar separation prior to charge exchange, a multi-harmonic buncher and a 12-MHz RFQ (Radio-Frequency Quadrupole) is proposed to supplement the present 300-kV injection system for the 25-MV tandem electrostatic accelerator. The RFQ will be mounted on a variable high voltage platform to accelerate ions with masses from 10 to 150 amu.

## 1 INTRODUCTION

The HRIBF [1] has been successfully conducting a research program of nuclear structure physics [2], nuclear reaction physics [3], and nuclear astrophysics [4] for several years. It is one of the first-generation Isotope Separation on Line (ISOL) RIB facilities. At the HRIBF, the target/ion source is mounted on a 300-kV rated platform. A light-ion beam from the Oak Ridge Isochronous Cyclotron (ORIC) bombards the target producing radioactive nuclei that are extracted from the ion source as 20-keV singly charged positive ions and subsequently accelerated to 40 keV. Negative ions are required for injection into the 25-MV tandem electrostatic accelerator. Therefore, after first-stage magnetic mass separation, the 40-keV singly charged positive-ion beam is injected into a recirculating jet cesium vapor charge-exchange cell. The energy spread of the resulting negative-ion beam is several tens of eV for an injected positive atomic beam and several hundreds of eV for an injected positive molecular beam, the increase being due to the three-body kinematics of molecular breakup. The negative-ion beam is finally accelerated to a nominal energy of 200 keV and sent through a second stage magnetic isobar separator with an intrinsic mass resolution of 20,000 prior to injection into the tandem accelerator.

Unfortunately, the actual mass resolution of the isobar separator is significantly degraded by the energy spread due to positive atomic beams and essentially eliminated by the energy spread due to positive molecular beams. This is a serious problem, particularly in the case of the high isobar contamination of RIBs that are produced from fission products. Various element specific isobar separation techniques are available [5], but magnetic isobar separation is applicable to a wide range of RIBs. A multi-harmonic buncher and a 12-MHz RFQ is proposed here to supplement the existing high voltage platform system. Charge exchange would be performed after isobar separation and before injection into the RFQ, which, in turn, would inject into the tandem accelerator.

A negative-ion beam cooler is also being studied at ORNL to reduce both the longitudinal and transverse emittance of RIBs [6]. If successful, the cooler could be used with the 300-kV platform system when negative-ion sources are used and charge exchange is not required.

## 2 LAYOUT OF INJECTION SYSTEM

Fig. 1 shows the present injection system for the HRIBF. Fig. 2 shows the proposed injection system for the HRIBF consisting of a multi-harmonic buncher and a 12 MHz RFQ. In the new system, the RIB is transmitted to the isobar separator at 60 keV instead of 200 keV. The charge-exchange cell has been moved downstream of the isobar separator. The RFQ is mounted on a $\pm 100$-kV platform to accelerate ions with masses from 10 to 150 amu. Due to severe space restrictions, both the diameter and the length of the RFQ are limited to approximately one meter. Consequently, there is no adiabatic bunching section, which would result in an RFQ more than three meters long. A multi-harmonic buncher that is designed to capture more than 80% of the injected DC beam can be located at ground potential within the available space.

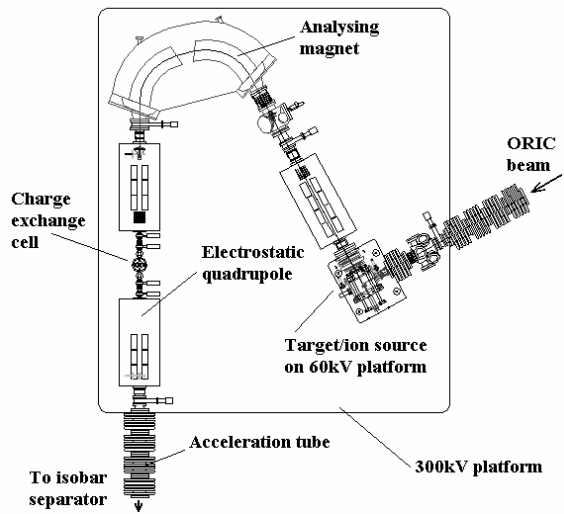

Fig. 1. Present injection system for the HRIBF.

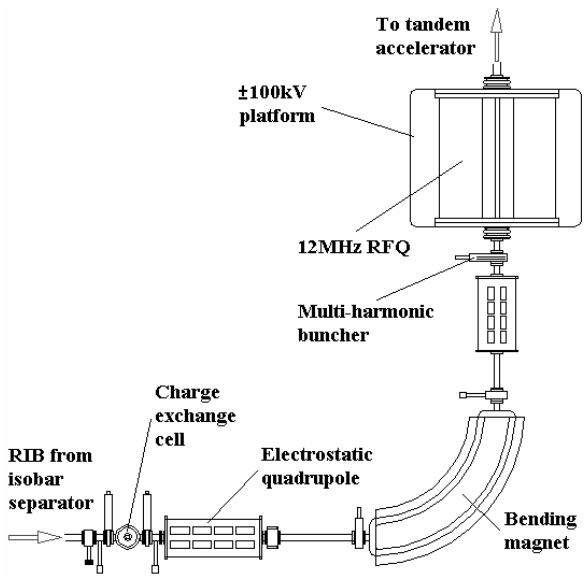

Fig. 2. Proposed RFQ injection system for the HRIBF.

In the RFQ injection system, the target/ion source is at 60 kV, and the first-stage mass separator and charge-exchange cell are now at ground potential. This will enhance operation and maintenance of the proposed injection system. Fig. 3 shows the beam envelopes from the image slit of the isobar separator to the entrance of the RFQ, calculated with the TRACE3D code [7].

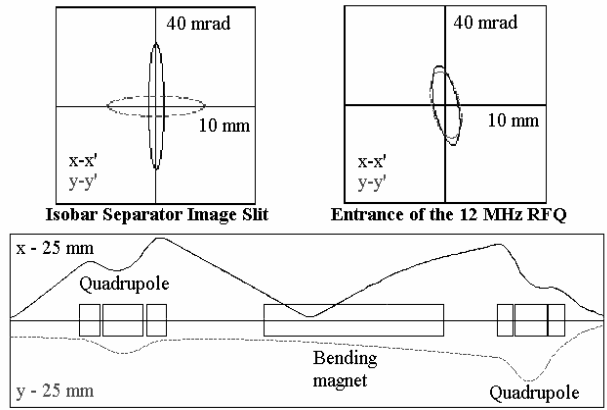

Fig. 3. Beam envelopes of the RFQ injection system from the isobar separator image slit to the entrance of the RFQ.

## 3 MULTI-HARMONIC BUNCHER

The multi-harmonic buncher will have 12, 24, 36, and 48 MHz modes with relative voltage amplitudes of 1.0, 0.44, 0.23, and 0.12, respectively. Since the beam may have several hundred eV energy spread after charge exchange, the buncher should be as close as possible to the RFQ to minimize time dispersion. Mounting the buncher on the ±100-kV platform along with the RFQ would minimize the distance between the two at the expense of requiring more RF power at high voltage. Table 1 lists features of a buncher located at ground potential for different ion masses. Since the time spreads of the bunched beams are less than the acceptance of the RFQ, this buncher design is satisfactory.

Table 1. Features of the multi-harmonic buncher.

| Ion mass | Equivalent drift length (cm) | Bunching voltage (kV) | Full width (ns) |
|---|---|---|---|
| 10 | 222.2 | 0.695 | 12.0 |
| 20 | 97.9 | 1.11 | 8.7 |
| 40 | 50.8 | 1.55 | 7.4 |
| 80 | 32.2 | 1.81 | 7.7 |
| 150 | 25.0 | 1.82 | 10.1 |

The longitudinal phase space for a 20-amu bunched beam and the longitudinal acceptance separatrix of the 12-MHz RFQ are shown in Fig. 4. The buncher can capture 84% of the DC beam for injection into the RFQ in principle. However, there is a 15% beam loss due to the buncher grids [8] that reduces the overall efficiency to approximately 70%. This is less than that of an RFQ with an adiabatic bunching section, but still acceptable.

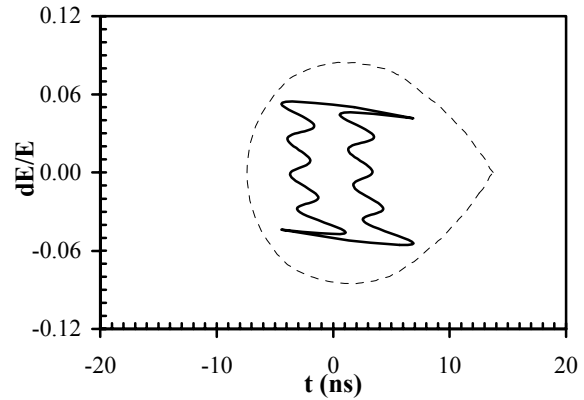

Fig. 4. Longitudinal phase space of a 20-amu bunched beam and the acceptance separatrix of the 12-MHz RFQ.

## 4 HEAVY ION RFQ

A 12-MHz heavy-ion RFQ with overall length of one meter has been designed using the code PARMTEQ [9]. Fig. 5 shows the various beam profiles at the exit of the RFQ with an injected 1-keV/amu 150-amu singly charged negative beam. The normalized transverse acceptance of the RFQ is 0.6 $\pi$·mm·mrad. The maximum required RF input power is approximately 15 kW as with the Atlas 12 MHz RFQ [10]. Because the RFQ has a narrow velocity acceptance, it must be mounted on a ±100-kV platform in order to accelerate RIBs from 10 to 150 amu. The injection parameters for different ion masses are listed in Table 2.

Various existing and proposed RFQs utilize split-coaxial [11,12], ring-connected [13], split ring [14] or four-rod spiral [15] structures to achieve low resonate frequencies. Since the proposed HRIBF RFQ has 94 cm electrodes, a split-coaxial structure would exceed the one-meter tank diameter restriction. However, a four-vane spiral structure with its larger end capacitances should reduce the tank diameter sufficiently. Fig. 6 shows the electrode structure of the RFQ. Some design parameters for the HRIBF RFQ are given in Table 3. Further mechanical, thermal, and RF efficiency studies need to be conducted.

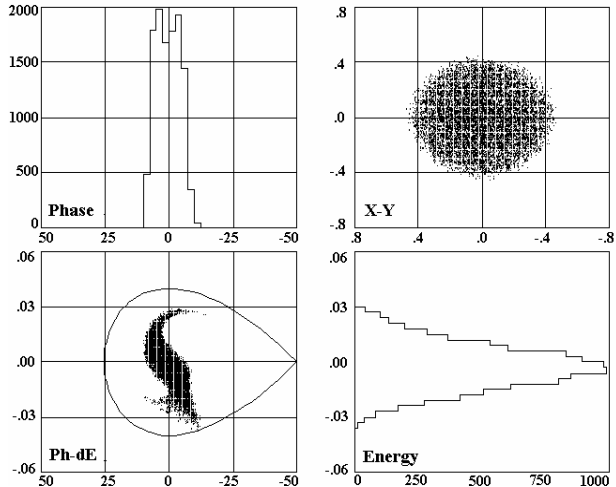

Fig. 5. PARMTEQ simulation of beam profiles at exit of the 12-MHz RFQ (Units: energy in MeV, phase in degree, x-y in cm). Ion mass: 150 amu; Charge: -1; Injection energy: 150 keV; Output energy: 774 keV; Normalized transverse acceptance: 0.6 π·mm·mrad.

Table 2. Injection parameters for different ion masses.

| Mass | $E_{IN}$ keV | $U_{PLATFORM}$ kV | $U_{RFQ}$ kV | $E_{OUT}$ keV |
|---|---|---|---|---|
| 10 | 60 | -50 | 6.67 | 102 |
| 20 | 60 | -40 | 13.3 | 143 |
| 40 | 60 | -20 | 26.7 | 226 |
| 60 | 60 | 0 | 40.0 | 310 |
| 80 | 60 | 20 | 53.3 | 393 |
| 100 | 60 | 40 | 66.7 | 476 |
| 120 | 60 | 60 | 80.0 | 559 |
| 140 | 60 | 80 | 93.3 | 642 |
| 150 | 60 | 90 | 100 | 684 |

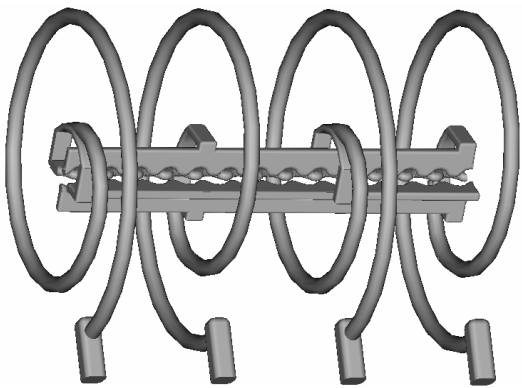

Fig. 6. Electrode structure of a four-vane spiral RFQ.

Table 3. Design parameters for the HRIBF RFQ.

| | |
|---|---|
| Resonate frequency | 12 MHz |
| Injection energy | 1 keV/amu |
| Output energy | 5 keV/amu |
| Inter vane voltage | 100 kV |
| Vane length | 94.2 cm |
| Number of cells | 33 |
| a | 0.8 cm |
| m | 1.5 |
| $\phi_S$ | -25° |

## 5 CONCLUSIONS

A multi-harmonic buncher and 12-MHz heavy ion RFQ injection system can allow magnetic isobar separation prior to charge exchange of positive atomic and molecular RIBs at the HRIBF. Isobar contamination of RIBs will be significantly reduced with the proposed injection system. Operation and maintenance of the target/ion source and other devices presently mounted on the 300-kV platform will be simplified.

## ACKNOWLEDGEMENT

Research at Oak Ridge National Laboratory is supported by the United States Department of Energy under contract DE-AC05-00OR22725 with UT-Battelle, LLC.

## REFERENCES

1. http://www.phy.ornl.gov/hribf.
2. D.C. Radford, et al., Phys. Rev. Lett. 88 (2002) 222501-1.
3. J. Gomez del Campo, et al., Phys. Rev. Lett. 86 (2001) 43.
4. D. W. Bardayan et al., Phys. Rev. Lett. 83 (1999) 45.
5. P. Van Duppen, A. Andreyev, et al., Nucl. Instrum. Methods B 134 (1998) 267-270.
6. Y. Liu, J. F. Liang, G. D. Alton, et al., Proc. PAC 2001, pp. 3885-3887, Chicago, USA, June 2001.
7. TRACE3D, Los Alamos Accelerator Code Group, LANL.
8. F. J. Lynch, R. N. Lewis, et al., Nucl. Instrum. Methods 159 (1979) 245-263.
9. RFQ Design Codes, Los Alamos Accelerator Code Group, LANL.
10. M. P. Kelly, P. N. Ostroumov, et al., Proc. PAC 2001, pp. 506-508, Chicago, USA, June 2001.
11. R. A. Kaye, K. W. Shepard, et al., AIP Conf. Proc. 473 (1998) pp. 528-535.
12. S. Arai, Y. Arakaki, et al., Proc. LINAC 96, WE203 (http://linac96.web.cern.ch/Linac96/) August 1996.
13. V. A. Andreev, A. A. Kolomiets, et al., Proc. PAC 1997, pp. 1090-1092, Vancouver, Canada.
14. R. Laxdal, R. Baartman, et al., Proc. LINAC 98, pp. 786-788, Chicago, USA, August 1998.
15. A. Schempp, Nucl. Instrum. Methods A 278 (1989) 169.